\documentclass[12pt]{spieman}  
\usepackage{amsmath,amsfonts,amssymb}
\usepackage{graphicx}
\usepackage{setspace}
\usepackage{tocloft}

\title{Physics-informed neural networks for diffraction tomography}

\author[a*]{Amirhossein Saba}
\author[a \dag]{Carlo Gigli}
\author[a \dag]{Ahmed B. Ayoub}
\author[a]{Demetri Psaltis}
\affil[a]{Optics Laboratory, École polytechnique fédérale de Lausanne, CH-1015, Lausanne, Switzerland}

\cftpagenumbersoff{figure}
\cftpagenumbersoff{table} 
\begin{document} 
\maketitle

\begin{abstract}
We propose a physics-informed neural network as the forward model for tomographic reconstructions of biological samples. We demonstrate that by training this network with the Helmholtz equation as a physical loss, we can predict the scattered field accurately. It will be shown that a pretrained network can be fine-tuned for different samples and used for solving the scattering problem much faster than other numerical solutions. We evaluate our methodology with numerical and experimental results. Our physics-informed neural networks can be generalized for any forward and inverse scattering problem.
\end{abstract}


{\noindent \footnotesize\textbf{*} Corresponding author: Amirhossein Saba,  \linkable{amirhossein.sabashirvan@epfl.ch} }

{\noindent \footnotesize\textbf{\dag} These authors contributed equally to this work.}

\begin{spacing}{2}   

\section{Introduction}
\label{sect:intro}  
Optical diffraction tomography (ODT) is an imaging technique for extracting the three-dimensional (3D) refractive index distribution of a sample, for example, a biological cell using multiple two-dimensional (2D) images acquired at different illumination angles. The refractive index of the sample provides useful morphological information, making ODT an interesting approach for biological applications \cite{choi2007tomographic,sung2009optical,jin2017tomographic}. The conventional method to reconstruct the 3D refractive index from multiple projections was proposed by Emil Wolf in 1969 \cite{wolf1969three}. In Wolf’s method, the 3D Fourier domain of the refractive index is filled with 2D Fourier transforms of the measured scattered fields. However, due to the finite number of projections, limited numerical aperture (NA) of the optical system, and the single scattering approximation, missing frequencies are causing some distortion, elongation, and underestimation of the reconstructed refractive index \cite{lim2015comparative}.

In the last years, many different iterative methods have been proposed to reconstruct accurate refractive indices from ill-posed measurements \cite{lim2015comparative,kamilov2015learning,chowdhury2019high,lim2019high,pham2020three}. The main idea behind these iterative approaches is to use a forward model which predicts the candidate projections for the current estimation of the refractive index in that iteration, compare this 2D prediction with the experimental measurements of the projection as a loss function and update the estimation of the refractive index upon the minimization of this loss function plus possibly any other prior knowledge, e.g. sparsity conditions. Importantly, such an iterative scheme requires an analytical/semi-analytical model in order to backpropagate the computed loss and update the estimation of the refractive index. This precludes the use of common mesh-based numerical solvers like finite difference and finite element methods. In Ref.~\citenum{lim2015comparative} the authors use a linear (single-scattering) forward model, in the approach proposed in Refs.~\citenum{kamilov2015learning,chowdhury2019high,lim2019high}, referred as learning tomography, the forward model is beam propagation method (BPM), and in Ref.~\citenum{pham2020three} the authors resort to the Lippmann–Schwinger equation. The forward models used in these iterative solutions either have inaccuracies in the cases of multiple-scattering and high-contrast samples, or they are computationally demanding. As a result, presenting a fast, accurate, and differentiable forward model is necessary to be used in iterative ODT. Physics-informed neural networks (PINNs), can be a good candidate for solving forward scattering problem and being used in iterative tomographic reconstruction.

PINNs have recently gotten intensive research attention for solving different complex problems in physics \cite{karniadakis2021physics,cai2022physics}. These networks use physics laws as the loss function instead of the data-driven loss functions. In conventional supervised deep learning, a large dataset of labeled examples is used for the training process: by comparing the known ground truth with the predictions from a deep multi-layer neural network, one can construct a loss function and tune the parameters of the network in order to solve complex physical problems. Different examples of these data-driven neural networks are proposed for optical applications such as resolution enhancement \cite{rivenson2017deep}, imaging through multi-mode fibers \cite{borhani2018learning,rahmani2018multimode}, phase retrieval \cite{rivenson2018phase}, and ODT \cite{lim2020three}. In these networks, the knowledge acquired by the network strongly depends on the statistical information provided in the dataset, and training such a network requires access to a large dataset. In contrast, PINNs directly minimize the physical residual from the corresponding partial differential equation (PDE) that governs the problem instead of extrapolating physical laws after going through a large amount of examples. In the pioneering approach proposed by Lagaris et al. \cite{lagaris1998artificial}, the neural network maps independent variables, such as spatial and time coordinates, to a surrogate solution of a PDE. By applying the chain rule, for example through auto-differentiation integrated in many deep-learning packages, one can easily extract the derivatives of the output fields with respect to the input coordinates and consequently construct a physics based loss \cite{Lu2021}. The correct prediction can be therefore retrieved by minimizing the loss with respect to the network weights. This approach has been used to solve nonlinear differential equations \cite{raissi2019physics,hashemi2019deep,mao2020physics,jin2021nsfnets}, to realize the forward model in the inverse design of optical components \cite{chen2020physics}, and to extract material parameters in near field microscopy \cite{chen2022}.

A different idea was proposed recently in \cite{lim2022maxwellnet}  to solve Maxwell's equations for a set of examples with different permittivity distributions. The calculation of physical loss, in this case, is based on the finite difference scheme, and in contrast to the previous approach which is trained for a specific example, this model proved to be well-suited for cases in which fast inference is required. In this paper, we use a similar approach, able to solve different forward scattering problems, such as light scattering from biological cells. Then we demonstrate using our PINN as a forward model in an
iterative reconstruction of the 3D distribution of the refractive index.

\section{Methodology}

The main idea of our work, summarized in Fig.~\ref{fig:sketch}, consists of two blocks. The first, MaxwellNet, is a neural network which takes as an input the refractive index distribution $n(\mathbf{r})$ and predicts the scattered field $U^s$. Its structure is based on the U-Net architecture \cite{ronneberger2015u}, and the training is performed on a large dataset of digital phantoms using a physics-defined loss function. Then, this network is used as a forward model in a second optimization task which compares the fields predicted by MaxwellNet for a candidate RI distribution with the ground-truth projections, e.g. computed numerically or evaluated experimentally, and updates $n(\mathbf{r})$ up to convergence.

\subsection{Forward Model: MaxwellNet}

In this subsection, we describe the implementation of a PINN which predicts the scattered field for a known input RI distribution. For the sake of simplicity, we first describe the method for the 2D case, but we will show the extension to 3D in the following. In this case, MaxwellNet takes as an input the RI distribution as a discrete array of shape  ${N_x{\times}N_z}{\times}1$ and we do expect an output with size ${N_x{\times}N_z}{\times}2$, where the two channels correspond to the real and imaginary parts of the complex field. Among all the available architectures, the choice of U-Net appears favorable as we do expect to embed the latent features of the RI distributions in a lower dimensional space through consecutive 2D convolutions and then retrieve the complex electromagnetic field in the same spatial points through the decoding step. A similar architecture was also proven to provide good accuracy for the computation of the scattered field from micro lenses \cite{lim2022maxwellnet}. We implement the present network in TensorFlow 2.6.0. For each step in the encoder, we use two Conv2D layers, each followed by batch-normalization and \textit{elu} activation function. A total number of five layers is adopted to encode the information and each one is terminated with average pooling to reduce the dimension. The maximum number of channels that we get in the latent space is 512. On the decoder side, we used transposed convolutional layers to the output with the size ${N_x{\times}N_z}{\times}2$ (or ${N_x{\times}N_y{\times}N_z}{\times}2$ in the 3D case). It should be noted that we also use residual skip connections from the encoder branch. In common data-driven training, we would tune the weights of this network by minimizing the difference between predictions and ground-truth data computed with numerical solvers, in turn requiring a large database of simulations and consequently a massive computational cost. Here we do not provide input-output pairs, instead we train the network by requiring that the Helmholtz equation is satisfied on the predicted field. In order to speed up the training and improve performances, we require the network to predict the slowly varying envelope of the scattered field $U^{s}_{env}$ being the scattered field obtained after demodulated by the fast-oscillating component along propagation direction $U^s = U^{s}_{env}e^{jk_{0}n_0z}$. We define a physics-informed loss function to be minimized by updating the weights of the network:

\begin{figure}
\begin{center}
\begin{tabular}{c}
\includegraphics[height=6.5cm]{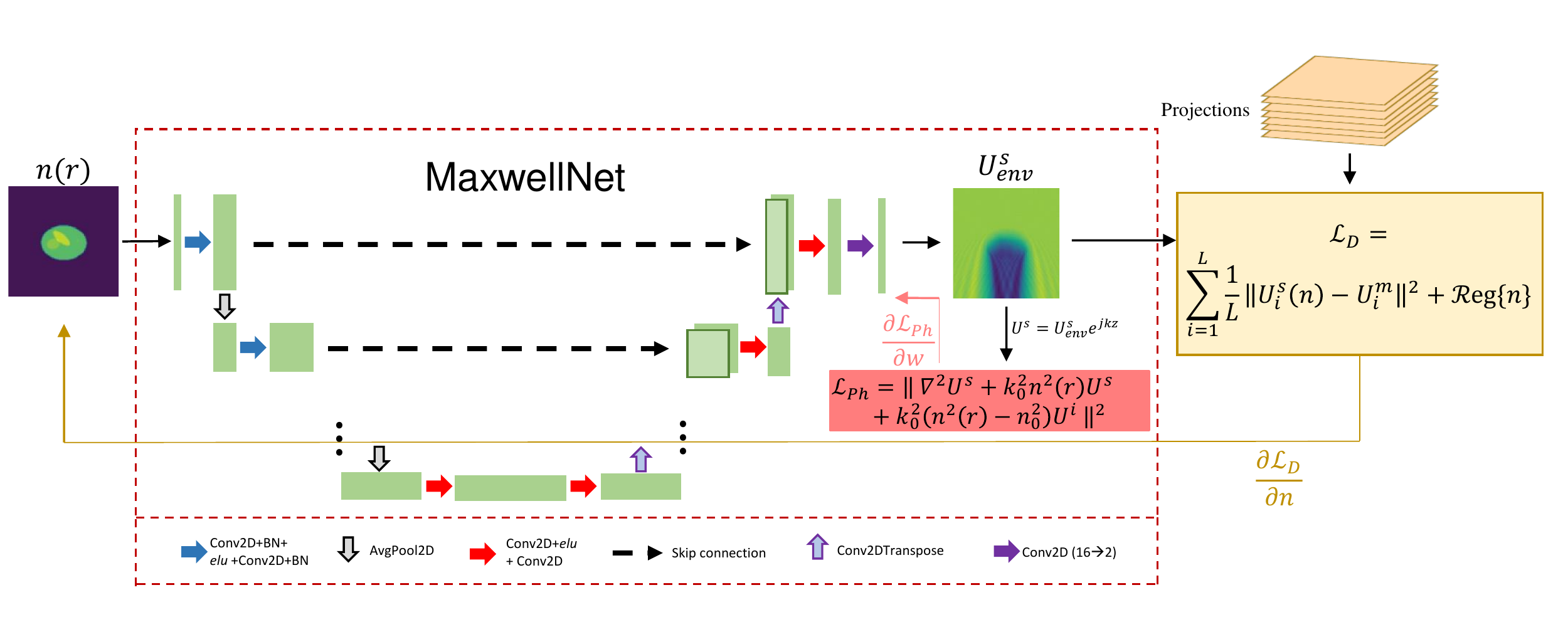}
\end{tabular}
\end{center}
\caption {Schematic description of MaxwellNet, with U-Net architecture, and its application for tomographic reconstruction. The input is a refractive index distribution and the output the envelope of the scattered field. The output is modulated by the fast-oscillating term $e^{jk_0n_0z}$ to compute the physics-informed loss for tuning the weights. For tomographic reconstruction, we minimize a data-driven loss based on the difference between measured and calculated projections using MaxwellNet. A regularization term can be added to improve the reconstruction.}
\label{fig:sketch}
\end{figure} 

\begin{equation}
    \mathcal{L}_{Ph} = \sum_{r}^{} \frac{1}{N}\left \|\{{\nabla}^2+{k_0^2}{n^2(r)}\}U^s+ {k_0^2}(n^2(r)-n_0^2)U^i  \right \|^2
    \label{eqn:physicalloss}
\end{equation}
where, $k_0$ is the wave-number which is $k_0=2\pi/\lambda$ and $\lambda=1.030 \mu m$ is the wavelength. $n(r)$ is the RI distribution and $n_0$ is the RI of the background medium. The summation in Eq.~\eqref{eqn:physicalloss} is done over the pixels of the computational domain and $N$ is the number of pixels. In order to implement the Laplacian in Eq.~\eqref{eqn:physicalloss}, we follow the Yee grid finite difference scheme, computing the derivative of variables by 2D convolutions with a $5\times5$ kernel \cite{yee1966numerical}. Additionally, light scattering is by definition an open boundary problem. In order to satisfy Sommerfeld radiation condition and confine the problem in a finite space we use a stretched-coordinate perfectly-matched layer (PML) \cite{chew19943d} at the edges of the simulation domain by introducing a complex coordinates transformation ($x\rightarrow x+if(x)$) when calculating the derivatives inside the PML region. We use the gradient of the so-computed physical loss function to update the weights of the neural network, $w$ through Adam optimizer:
\begin{equation}
    w \rightarrow w -\gamma_{Ph} \frac{\partial {\mathcal{L}_{Ph}}}{\partial w}
\end{equation}
When we train MaxwellNet for a class of samples, it can accurately calculate the field for unseen samples from the same class. However, the key point to mention is that if we want to use MaxwellNet for a different set of RI distributions, we can fix some of the weights, and adjust only a part of the network for the new dataset, instead of re-starting the training from scratch. This process, referred in the following as fine-tuning, is much faster than the original training of MaxwellNet. We will elaborate and discuss this interesting feature in section~\ref{sec3}.

\subsection{Optical diffraction tomography using MaxwellNet}

Once MaxwellNet has been trained on a class of RI distributions, it can be used to rapidly backpropagate reconstruction errors with an approach similar to learning tomography \cite{kamilov2015learning}. Let us assume that we measure $L$ projections $U^m_i$, with $i=0,...,L$, from an unknown RI distribution $\bar{n}(x,z)$ for different rotational angles. From these data, we can reconstruct a first inaccurate candidate $n(x,z)$ through the Wolf's transform using Rytov approximation. By feeding MaxwellNet with $n(x,z)$ we predict the complex scattered fields $U^s_i$ for the same $L$ angles. Consequently we can construct a data-driven loss function $\mathcal{L}_{D}$ given by the difference $\|U^s_i-U^m_i\|$ plus any additional regularizer, compute its gradient through auto-differentiation, update $n(x,z)$ and iterate up to convergence:

\begin{equation}
    \mathcal{L}_{D} = \sum_{i=1}^{L}\frac{1}{L}{\left\|{U^s_i(n) - U^m_i}\right\|}^2+ \mathcal{R}\textrm{eg}\{n,U^s_i(n) \}
    \label{eqn:datadrivenloss}
\end{equation}

\begin{equation}
    n \rightarrow n -\gamma_{D} \frac{\partial {\mathcal{L}_{D}}}{\partial n}
\end{equation}

Also in this case, we use Adam optimizer for updating RI values. The regularizer in Eq.~\eqref{eqn:datadrivenloss} consists of three parts, a total-variation (TV), a non-negativity and a physics-informed terms, $\mathcal{R}\textrm{eg}\{n,U^s_l(n) \}=\lambda_{TV}\mathcal{R_{TV}}(n)+\lambda_{NN}\mathcal{R_{NN}}(n)+\lambda_{Ph}\mathcal{L}_{Ph}(n,U^s)$. The TV regularizer helps smoothing the RI reconstruction and the non-negativity regularizer adds the prior information that $n(x,z)$ should be larger than the background refractive index:
\begin{subequations}
\begin{equation}
\mathcal{R_{TV}}(n) = \sum_{r}^{}\sqrt{\left | \nabla_xn(r)) \right |^2+\left | \nabla_yn(r)) \right |^2+\left | \nabla_zn(r)) \right |^2}
\end{equation}
\begin{equation}
\mathcal{R_{NN}}(n) = \sum_{r}^{}{\mathrm{min}(n(r)-n_0,0)}^2
\end{equation}
\end{subequations}

Importantly, we have to remark that MaxwellNet is trained for a specific dataset and accurately predicts the scattered field for RI distributions that are not too far from this set. To take into account this effect we add the physics-informed loss to the regularizer. This further correction term helps to find RI values in a way that MaxwellNet can predict the scattered field for them correctly. In contrast to TV and non-negativity constraints that are used due to the ill-posedness of the ODT problem, the physics-informed regularizer is necessary in our methodology to ensure that the index distributions remain within the domain in which MaxwellNet has been trained. 

The key advantages of using MaxwellNet with respect to other forward models are three folds: differently from BPM, it can accurately calculate field scattering, considering reflection, multiple-scattering, or any other electromagnetic effects \cite{lim2015comparative,kamilov2015learning,chowdhury2019high,lim2019high}; once trained, the field computations are performed in milliseconds, much faster than Lipmann-Schwinger model; and finally, the data-driven error in Eq.~\eqref{eqn:datadrivenloss} can be easily backpropagated differently from commercially available full-vectorial solvers. We discuss the reconstruction results and compare them with other methods in subsection 3.2. 

\section{Results and Discussion}
\label{sec3}

\subsection{MaxwellNet results}
In this subsection, we evaluate the performance of MaxwellNet for the prediction of the scattered field from RI structures such as biological cells. At first, we check the performance on a 2D sample assuming that the system is invariant along the y axis. The number of pixels for our model are $N_x=N_z=256$ for both the x and z directions, and their size is $dx=100\mathrm{nm}$. We also use PML with the thickness of $1.6\mathrm{\mu m}$ at the edges of our computational domain. We create a dataset of digital cell phantoms and divide it into the training and testing sets. MaxwellNet has $\sim5.9\mathrm{M}$ parameters to train and we use the Adam optimizer with a learning rate of $1\times10^{-4}$ and batch training. The details about the dataset and training and validation losses are discussed in Appendix~\ref{sec:appB}. We train and test MaxwellNet in Tensorflow 2.6.0 on a desktop computer (Intel Core i7-9700K CPU, 3.6GHz, 64GB RAM, GPU GeForce RTX 2080Ti). 

\begin{figure}
\begin{center}
\begin{tabular}{c}
\includegraphics[height=10.2cm]{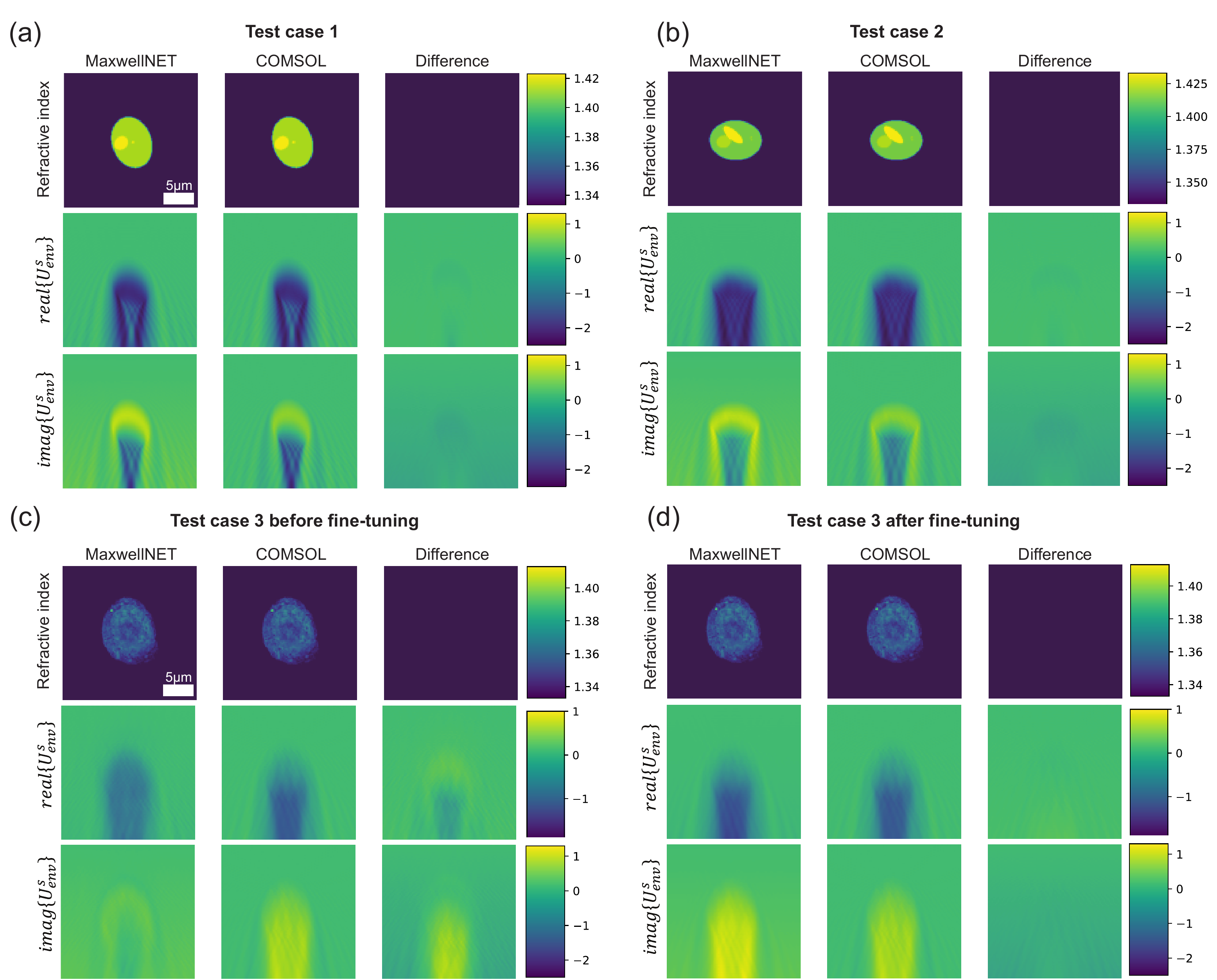}
\end{tabular}
\end{center}
\caption 
{Results of MaxwellNet and its comparison with COMSOL. (a,b) Two test cases from the digital phantom dataset and the prediction of the real and imaginary of the envelope of the scattered fields using MaxwellNet, COMSOL and their difference. (c) Scattered field predictions from the network trained in (a,b) for the case of an experimentally measured RI of HCT-116 cancer cell and comparison with COMSOL. The difference between the two is no longer negligible. (d) Comparison between MaxwellNet and COMSOL after fine-tuning the former for a set of HCT-116 cells. MaxwellNet predictions reproduces much more accurately results after fine-tuning.}
\label{fig:MaxwellNet_training}
\end{figure} 

In Fig.~\ref{fig:MaxwellNet_training}(a) and Fig.~\ref{fig:MaxwellNet_training}(b), we choose two random examples of the digital phantoms in the test set (which is not seen by the network during the training). For each test case, in the second and third rows, we present the prediction of the envelope of the scattered field by the network, and we compare it with the result achieved by the finite element method (FEM) using COMSOL Multiphysics 5.4. We can see a very small difference between the results of MaxwellNet and COMSOL, which we attribute to discretization error.  There are different schemes of discretization in two methods that can cause such differences. In order to quantitatively evaluate the performance of MaxwellNet, we define the relative error of MaxwellNet with respect to COMSOL as,
\begin{equation}
    error=\frac{\int{\left \| {{U}_{MaxwellNet}(r)-{U}_{COMSOL}(r)}\right \|^2dr}}{\int{\left \|{{U}_{COMSOL}(r)}\right \|^2dr}}
    \label{eqn:error}
\end{equation}
where ${U}_{MaxwellNet}$ and ${U}_{COMSOL}$ are the total fields calculated with MaxwellNet and COMSOL. The integration is done excluding the PML regions. The calculated relative errors for the test case 1 and the test case 2 in Fig.~\ref{fig:MaxwellNet_training} are $4.1\times10^{-2}$ and $4.6\times10^{-2}$, respectively.

It should be noted that once MaxwellNet is trained, the scattered field calculation is much faster than numerical techniques such as FEM. We present a time comparison in Table~\ref{table:table1}. For the test phantoms in Fig.~\ref{fig:MaxwellNet_training}, it took 17ms for MaxwellNet in comparison with 13s for COMSOL meaning three orders of magnitude acceleration.

\begin{figure}
\begin{center}
\begin{tabular}{c}
\includegraphics[width=14.1cm]{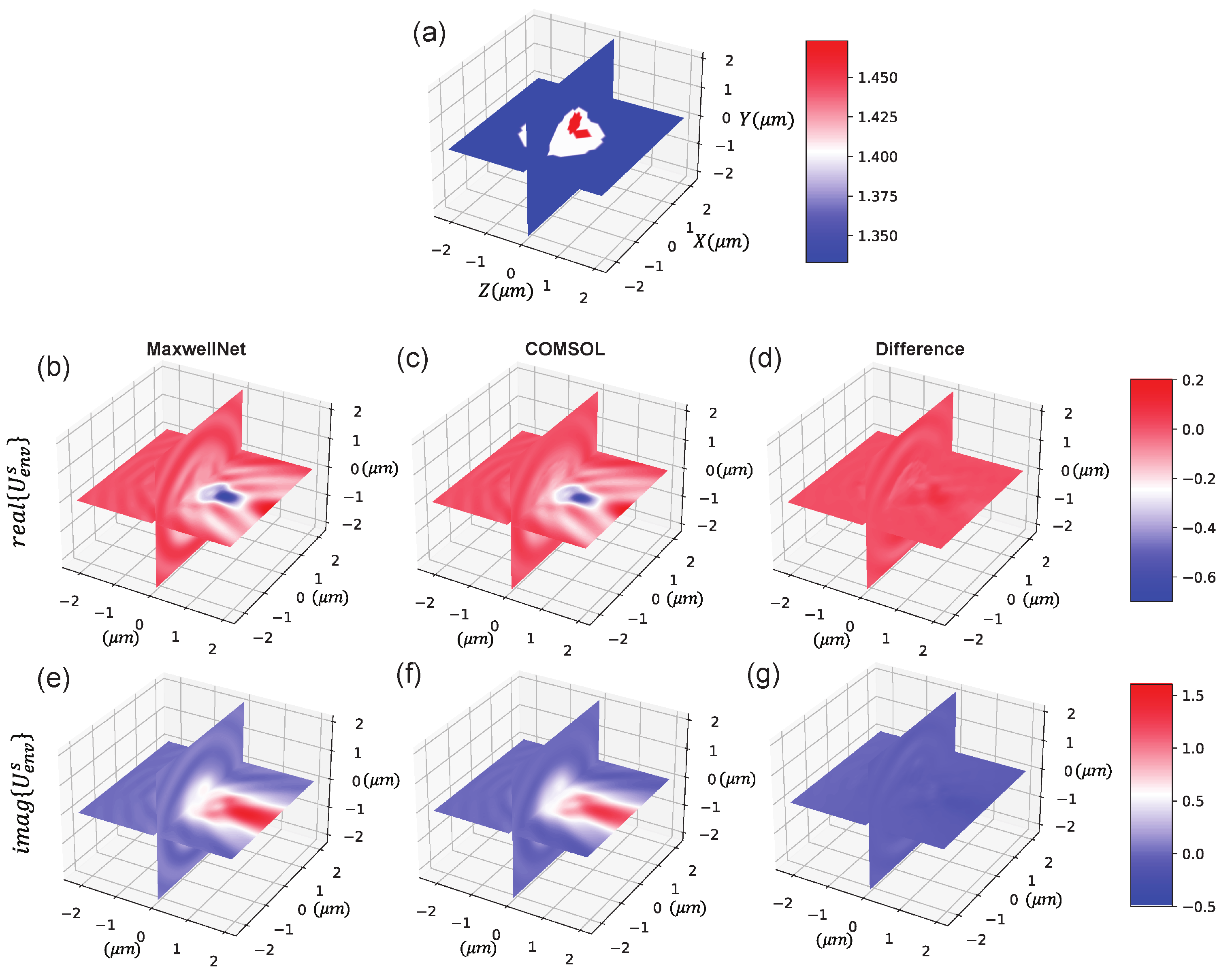}
\end{tabular}
\end{center}
\caption 
{Results of MaxwellNet3D and its comparison with COMSOL. The RI distribution is shown in (a). The real part of the envelope of the scattered field calculated by MaxwellNet3D is shown in (b), calculated by COMSOL in (c), and their difference in (d). The imaginary part of the envelope of the scattered field calculated by MaxwellNet3D, COMSOL, and their difference are presented in (e-g), respectively.} 
\label{fig:3dMaxwellnet}
\end{figure} 

Furthermore, performing a physics based instead of direct data-driven training holds promises for exploiting the advantages of Transfer learning\cite{tan2018survey}. Maxwell equations are general but having a neural network which predicts the scattered field for any class of RI distribution in milliseconds with a negligible physical loss is usually unfeasible. Most of the previous PINN studies for solving partial differential equations are trained for one example, and they will work for that specific example. In our case, U-Net architecture proved to be expressive enough to predict the field for a class of samples. However, if we use MaxwellNet for inference on a RI distribution completely uncorrelated with the training set, the accuracy drops. In order to evaluate MaxwellNet extrapolation capability, we considered the model trained on phantoms samples in Fig.~\ref{fig:MaxwellNet_training} and use it for inference on HCT-116 cancer cells. The comparison between MaxwellNet and COMSOL is shown in Fig.~\ref{fig:MaxwellNet_training}(c). The input of the network is a 2D slice of the experimentally-measured HCT-116 cell in the plane of best focus. The discrepancy between MaxwellNet and COMSOL is due to the fact that the former does not see examples of such RI distributions during the training. As a result, if we require accurate results for a new set of samples with different features, we have to re-train MaxwellNet for the new dataset, which would take a long time as reported in Table~\ref{table:table1}. However, it turns out that learning a physical law, as Maxwell equations, even though on a finite dataset is better suited than data-driven training for transfer learning on new batches. Indeed, we can use the pretrained MaxwellNet on digital phantoms and fine-tune some parts of the network for HCT cells achieving good convergence in a few epochs. In this example, we create a dataset of 136 RI distributions of HCT-116 cancer cells and divide them into the training and validation sets. Some examples of HCT-116 refractive index dataset are shown in Appendix~\ref{sec:appB}. A wide range of cells with different shapes are included in the dataset. We have single cancer cells, like Fig.~\ref{fig:MaxwellNet_training}(c), examples of cells in the mitosis process, or examples with multiple cells. In this case, we freeze the weights of the encoder part and fine-tune the decoder with the new dataset. We can see in Fig.~\ref{fig:MaxwellNet_training}(d) that after this correction step, the calculated field is much more accurate. As it can be seen in Table~\ref{table:table1}, the fine-tuning process is two orders of magnitude faster than a complete training from scratch.

The 2D case is helpful for demonstrating the method and rapidly evaluate the performances. Nevertheless, full 3D fields are required for many practical applications. We can straightforwardly recast MaxwellNet in 3D using arrays of size $N_x \times N_y \times N_z\times1$ as inputs of the network and requiring a $N_x\times N_y\times N_z\times2$ output, with the two channels corresponding to the real and imaginary of the envelope of the scattered field. In this case, the network consists of Conv3D, AveragePool3D, and Conv3DTranspose layers instead of 2D counterparts. As a benchmark test, we created a dataset of 3D phantoms with 200 examples (180 for training and 20 for testing). The computational domain is defined with $N_x=N_y=N_z=64$, $dx=100\mathrm{nm}$, and PML thickness of $0.8\mathrm{\mu m}$.

\begin{table}[b]
\caption{Computation time comparison} 
\label{tab:fonts}
\begin{center}       
\begin{tabular}{|c|c|c|c|} 
\hline
\rule[-1ex]{0pt}{3.5ex}  Dataset & 2D Phantoms & 2D HCT-116 & 3D Phantoms  \\
\hline\hline
\rule[-1ex]{0pt}{3.5ex}  MaxwellNet training/fine-tuning & 30.5h & 0.18h& 15.5h   \\
\hline
\rule[-1ex]{0pt}{3.5ex}  MaxwellNet inference&17.0ms&17.0ms&44.9ms \\
\hline
\rule[-1ex]{0pt}{3.5ex}  COMSOL&13s&13s&2472s \\
\hline 
\end{tabular}
\end{center}
\label{table:table1}
\end{table} 

The 3D version of MaxwellNet has $\sim17.2\mathrm{M}$ parameters. We use Adam optimizer (learning rate $=1\times 10^{-4}$), and a batch size of 10. The results of the predicted field for an unseen example and its comparison with COMSOL are shown in Fig.~\ref{fig:3dMaxwellnet}. We can see that MaxwellNet performs as good as COMSOL in field calculation. The quantitative error described in Eq.~\ref{eqn:error} is $3.4\times10^{-3}$ for the 3D example of Fig.~\ref{fig:3dMaxwellnet}. There are some marginal differences due to different discretization schemes. However, we can see in Table~\ref{table:table1} that MaxwellNet is about 50000 faster than COMSOL in predicting fields (44.9 milliseconds versus 41.2 minutes). This result and the significant efficiency in the computation time highlight MaxwellNet potential for the calculation of the field in different applications. In the next subsection, we demonstrate how this method can be applied for improving ODT reconstruction fidelity.

\begin{figure}
\begin{center}
\begin{tabular}{c}
\includegraphics[height=4.12cm]{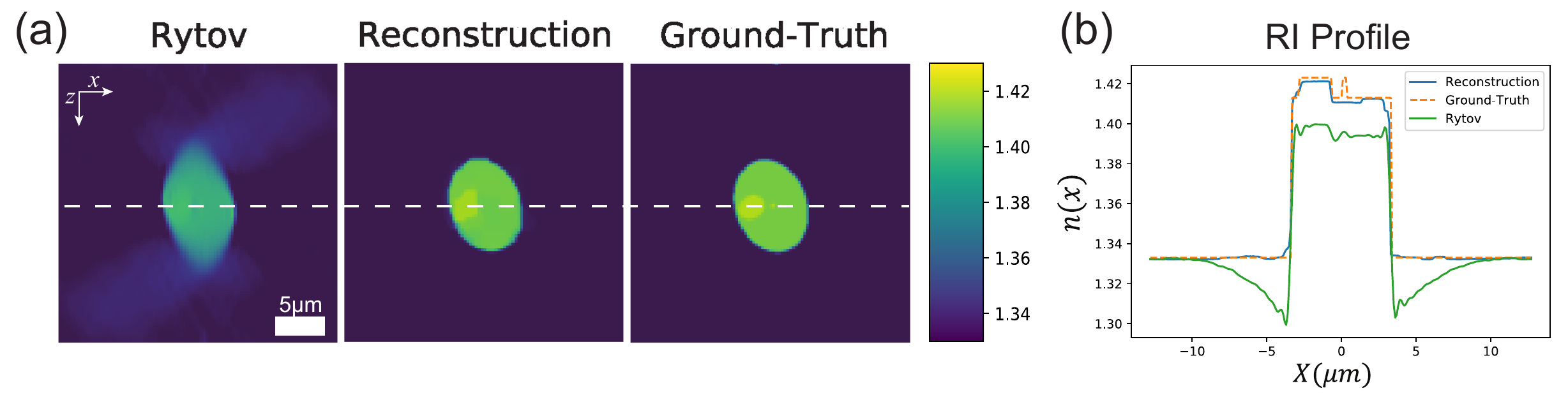}
\end{tabular}
\end{center}
\caption 
{Tomographic reconstruction of RI using MaxwellNet. (a) The RI reconstruction was achieved by Rytov, MaxwellNet, and the ground-truth. (b) 1D RI profile at $z=0$ (plane of best focus), for Rytov (green), MaxwellNet (blue), and the ground-truth (orange).} 
\label{fig:tomography}
\end{figure} 

\subsection{Tomographic reconstruction results}
To show the ability of MaxwellNet to be used for different imaging applications, we implement an optimization task with MaxwellNet as the forward model for ODT as explained in subsection 2.2. In this example, we consider one of the digital phantoms in the test set of Fig.~\ref{fig:MaxwellNet_training} and we use 2D MaxwellNet as the forward model to compute the 1D scattered field along the transverse direction $x$ for $N=81$ different rotation angles $\theta$. We restrict ourselves to the range $\theta\in[-40^{\circ},40^{\circ}]$ as this is consistent with the typical conditions in common tomographic setups. As is shown in Fig.~\ref{fig:tomography}(a), the Rytov reconstruction obtained from these field projections is elongated along z-axis and underestimated due to missing frequencies. We then minimize the loss function \eqref{eqn:datadrivenloss} to improve the RI reconstruction choosing as regularizer parameters $\lambda_{TV}=3.1\times10^{-7}$, $\lambda_{NN}=1\times10^{-1}$, $\lambda_{Ph}=5\times10^{-2}$ and Adam optimizer with initial learning rate of $3\times10^{-4}$. We also scheduled the learning rate, halving it every 1000 epochs to speed-up convergence. The resulting RI distribution after 3000 epochs is shown in Fig.~\ref{fig:tomography}. It can be seen that the reconstructed RI is not anymore underestimated nor elongated along the z-axis. This is a significant improvement in comparison with Rytov prediction. The missing details in the reconstructed RI, which can be better visible in the 1D cutline in Fig.~\ref{fig:tomography}(b), can be due to the missing information in 1D fields that the optimization of RI could not retrieve this information.

\begin{figure}[t]
\begin{center}
\begin{tabular}{c}
\includegraphics[height=11cm]{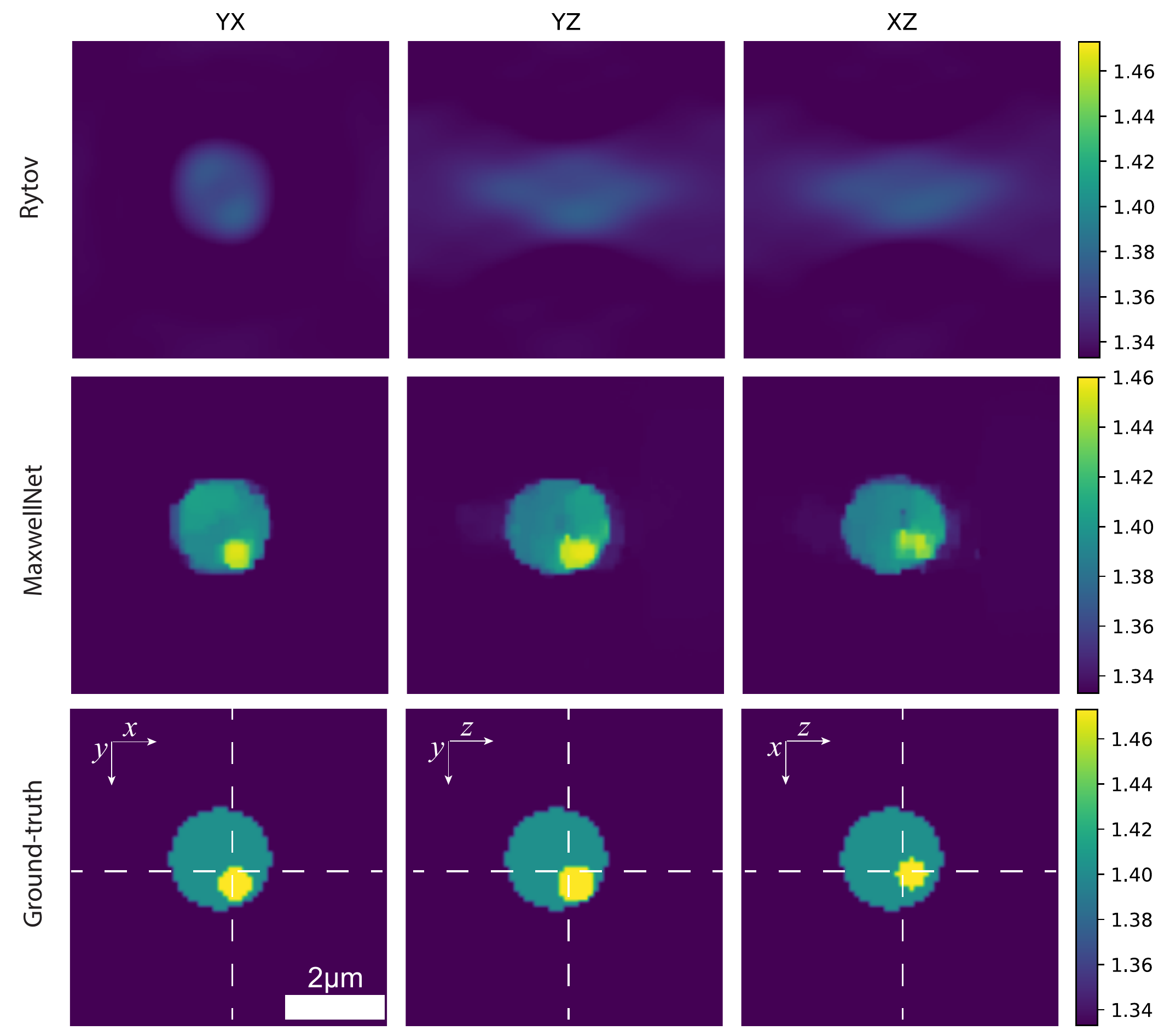}
\end{tabular}
\end{center}
\caption 
{Tomographic RI reconstruction of 3D sample using MaxwellNet. The RI reconstruction is achieved by Rytov, MaxwellNet, and the ground-truth in different rows at YX, YZ, and XZ planes in the center of the sample. A z-stack demonstration of the reconstruction is shown in Video 1 (Video 1, MP4, 1.4 MB).} 
\label{fig:tomography3D}
\end{figure} 

Next, we try a 3D digital phantom from the test set and we use 3D MaxwellNet as the forward model in our tomographic reconstruction method. Since generating synthetic data with COMSOL is time-consuming for multiple angles, we create synthetic scattered fields from the phantom with the Lippmann–Schwinger equation \cite{pham2020three}. We will show later an experimental example, where we illuminate the sample with a circular illumination pattern with an angle $\approx 10^{\circ}$. As a result, in this numerical example, we rotate the sample for 181 angles (including 1 normal incidence), equivalently to an illumination rotation with a fixed illumination angle of $10^{\circ}$. We keep the experimental conditions, $\lambda=1.030\mathrm{\mu m}$, and $n_0=1.33$. Then, we use these synthetic measurements for our optimization task along with TV, non-negativity, and physics-informed regularization. The reconstruction is achieved after 6000 epochs with $\lambda_{TV}=1.2\times10^{-8}$, $\lambda_{NN}=2\times10^{1}$, and $\lambda_{Ph}$ started with $5\times10^{-1}$ and divided by two every 500 epochs. The reconstructions are shown in Fig.~\ref{fig:tomography3D} in YX, YZ, and XZ planes. The first row shows the Rytov reconstruction where we can see a significant underestimation and elongation along z-axis which is due to the small illumination angle ($10^{\circ}$). The details in the reconstruction achieved using MaxwellNet are slightly blurred in comparison with the ground-truth as a result of low resolution with $\lambda=1.030\mathrm{\mu m}$.

\begin{figure}[t]
\begin{center}
\begin{tabular}{c}
\includegraphics[height=6.5cm]{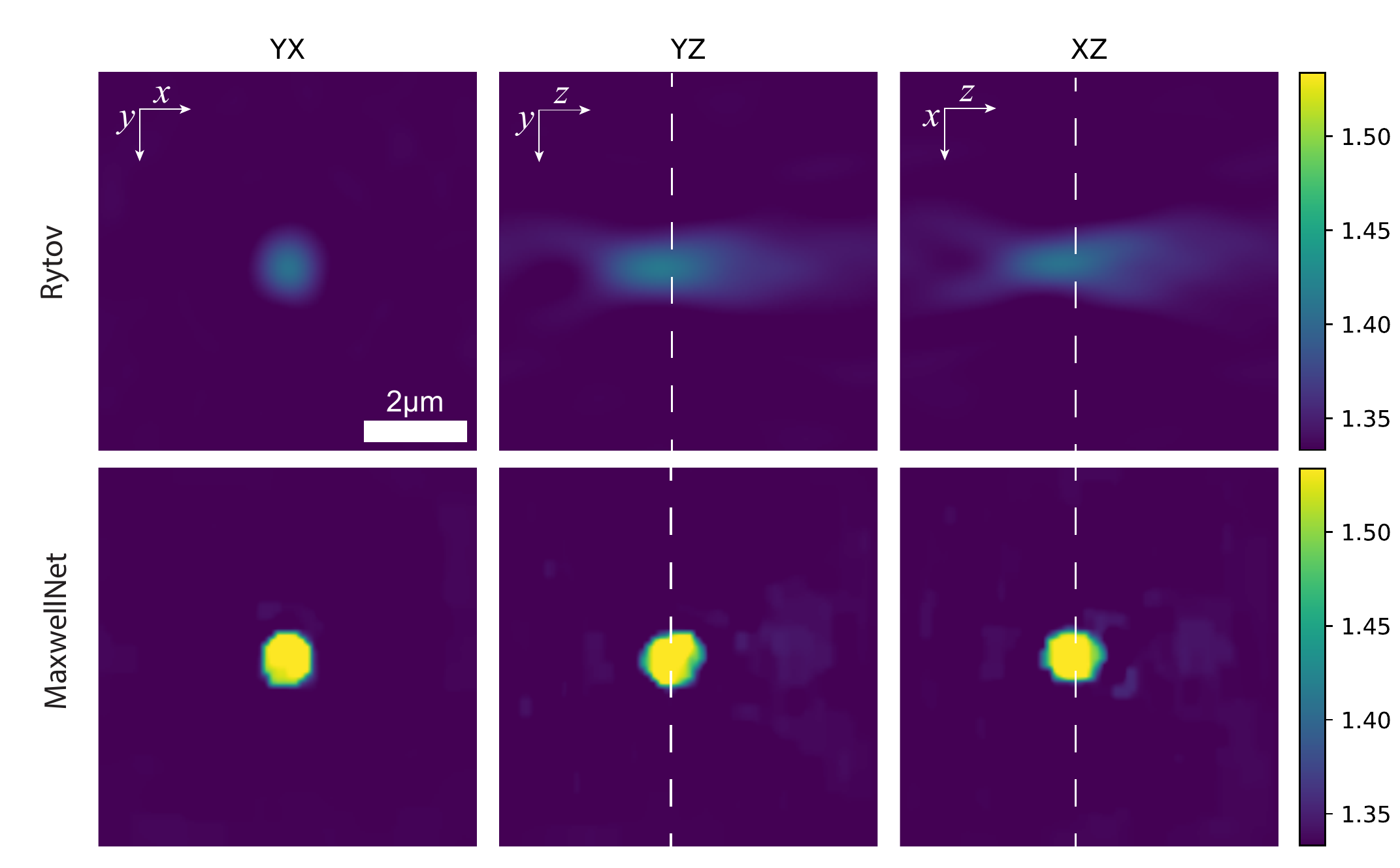}
\end{tabular}
\end{center}
\caption 
{Tomographic RI reconstruction of a polystyrene micro-sphere immersed in water. The projections are measured with off-axis holography for different angles. The RI reconstruction achieved by Rytov, and MaxwellNet are presented at YX, YZ, and XZ planes in the center of the sample.} 
\label{fig:tomographyexp}
\end{figure} 

We also evaluated our methodology experimentally. We mentioned earlier that MaxwellNet takes care of reflection as a forward model, and therefore, our reconstruction technique can be used for samples with high contrast. In our experimental analysis, we try a polystyrene micro-sphere immersed in water, where we expect to have a $\sim0.25$ refractive index contrast. Polystyrene micro-spheres (Polybead{\textregistered} Polystyrene 2.0 Micron) are immersed in water and placed between two \#1 glass coverslips.
We have an off-axis holographic setup where we use a yttrium-doped fiber laser (Amplitude Laser Satsuma) with $\lambda=1.030\mathrm{\mu m}$ and we change the illumination angle with two Galvo mirrors. Using a delay path, the optical length of the reference and signal arms are matched. We measure holograms for 181 illumination angles and extract the phase and amplitude of the complex scattered fields using Fourier holography. More details about the experimental setup are discussed in Appendix~\ref{sec:appC}. Then, we use the extracted scattered fields for different projections for our optimization task to reconstruct the 3D RI distribution of the sample. The experimental projections are 2D complex fields that are imaged in the center of the sample using a microscope objective lens and we can propagate them in the background medium to calculate the scattered field in any other plane, perpendicular to $z$-axis, after the sample. This 2D field can be compared with the output of MaxwellNet in that plane, as described in Eq.~\ref{eqn:datadrivenloss}. Additionally, the experimental projections are based on illumination rotation and we interpolate them to achieve the equivalent sample rotation projections. We iteratively optimize the loss function in Eq.~\ref{eqn:datadrivenloss} for 2000 epochs where we use the regularization parameters of $\lambda_{TV}=3.8\times10^{-9}$, $\lambda_{NN}=5\times10^{1}$, and $\lambda_{Ph}$ started with $1.5\times10^{-1}$ and divided by two in every 500 epochs. The reconstruction is shown in Fig.~\ref{fig:tomographyexp}. It can be seen that the underestimation and z-axis elongation in the Rytov reconstruction are remarkably improved.

\section{Conclusion}

In summary, we proposed a PINN which rapidly calculates the scattered field from inhomogeneous RI distributions such as biological cells. Our network is trained by minimizing a loss function based on Maxwell equations. We showed that the network can be trained for a set of samples and could predict the scattered field for unseen examples which are in the same class. As our PINN is not a data-driven neural network, it can be trained for different examples in different conditions. Even though the network is not efficiently extrapolating to classes which are statistically very different from the training data-set, we showed that by freezing the encoder weights and fine-tuning the decoder branch, one can get a new predictive model in a few minutes. We believe that this can be further used for changing wavelength, boundary condition, or other physical parameters. 

We used our PINN as a forward model in an optimization loop to retrieve the RI distribution from the scattered fields achieved by illuminating the sample from different directions, known as optical diffraction tomography. This example shows the ability of MaxwellNet to be used as an accurate forward model in optimization loops for inverse design or inverse scattering problems.

\appendix    

\section{Calculation of Physics-informed Loss}
\label{sec:appA}
During the training of MaxwellNet, we calculate at each epoch the loss function in Eq.~\ref{eqn:physicalloss} for the network output. In order to evaluate the Helmholtz equation residual, we should numerically compute the term $ \frac{\partial^2 U^s}{\partial x^2}+\frac{\partial^2 U^s}{\partial y^2}+\frac{\partial^2 U^s}{\partial z^2}$. In the previous PINN papers for solving PDEs \cite{lagaris1998artificial,Lu2021,chen2020physics,chen2022,raissi2019physics,mao2020physics,jin2021nsfnets}, the inputs of the network are the spatial coordinates $x$, $y$, $z$, and the derivatives with respect to these variables can be calculated using the chain rule. In this implementation, the weights of the network can be trained to minimize the loss function for a single refractive index distribution, $n(r)$ in Eq.~\ref{eqn:physicalloss}. In our approach, the PINN gets the refractive index, $n(r)$, on a uniform grid as the input and finds the field on the same grid which minimizes the loss function for that refractive index. The output of the network is the 3D array of the scattered field envelope, and we use a finite difference scheme to calculate the derivative of the field with respect to the coordinates:
\begin{equation}
    \frac{\partial U^s}{\partial x} = \frac{ U^s((i+1){\Delta}x,j{\Delta}y, k{\Delta}z) -U^s((i-1){\Delta}x, j{\Delta}y,k{\Delta}z)}{2{\Delta}x}
\end{equation}
in which $(i, j, k)$ are the pixel indices and ${\Delta}x$, ${\Delta}y$, ${\Delta}z$, are the pixel sizes  along the $x$, $y$, and $z$ axes. This way, we can calculate $\frac{\partial U^s}{\partial x}$ by convolving $U^s$ with a kernel of $\left [ -1/2,0,1/2 \right ]$ along the $x$ axis. When computing electromagnetic fields, since the curl of the electric field gives the magnetic field and vice versa, a smart technique to improve accuracy is to use two staggered grids for discretizing fields, commonly referred to as Yee scheme \cite{yee1966numerical}. In practice, this can be easily implemented through two shifted convolutional kernels for the two grids, $\left [ -1/2,1/2,0 \right]$ and $\left [ 0,-1/2,1/2 \right]$. 

In order to minimize the discretization error, one can use a smaller pixel size, ${\Delta}x$ or higher order approximations. Here, we use the fourth-order finite difference scheme \cite{Fathy2008} in which convolutional kernels of $\left [ 0, +1/24, -9/8, +9/8, -1/24 \right ]$ and $\left [+1/24, -9/8, +9/8, -1/24, 0 \right ]$ are used for the calculation of the derivatives in Eq.~\ref{eqn:physicalloss}.

\section{Training and Fine-tuning of MaxwellNet}
\label{sec:appB}
\begin{figure}[t]
\begin{center}
\begin{tabular}{c}
\includegraphics[height=4.30cm]{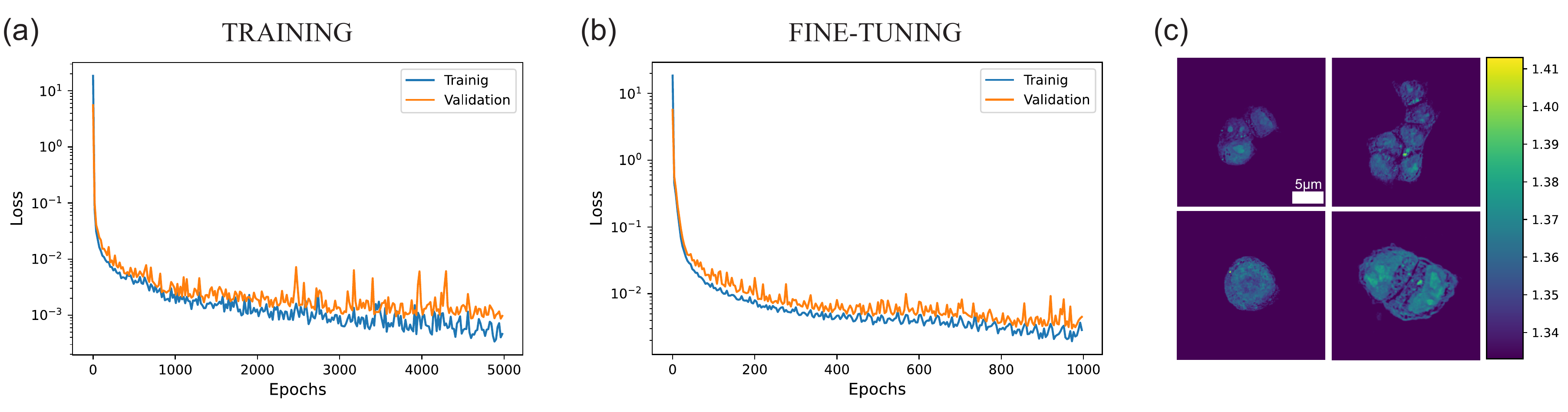}
\end{tabular}
\end{center}
\caption 
{Training and fine-tuning of MaxwellNet. (a) Training (blue) and validation (orange) loss of MaxwellNet for Digital cell phantoms dataset. (b) Fine-tuning the pretrained MaxwellNet for a dataset of HCT-116 cells for 1000 epochs. (c) Examples of the HCT-116 dataset.} 
\label{fig:Fig9}
\end{figure}

As mentioned in Section~\ref{sec3}, we create a dataset of digital cell phantoms to train and validate MaxwellNet. The dataset for 2D MaxwellNet includes 3000 phantoms with elliptical shapes oriented in different directions. The size of these phantoms is in the range of $5-10\mathrm{\mu m}$, their refractive index varies in the range of $(1.38,1.45)$, and the background refractive index is $n_0=1.33$. Two examples of these phantoms are shown in Fig.~\ref{fig:MaxwellNet_training}. We divide this dataset into
2700 phantoms for training and 300 phantoms for testing. We use batch training with a batch size of 10 for 5000 epochs. This training took 30.5 hours and after 5000 epochs, no significant decrease in the validation loss could be observed. The training and validation curves of the physical loss are presented in Fig.~\ref{fig:Fig9}(a). This figure shows that MaxwellNet performs very well for out-of-sample cases. 

We discussed in Section~\ref{sec3} using MaxwellNet which was trained for cell phantoms to predict the scattered field for real cells. A dataset of HCT-116 cancer cells is used for this purpose. The 3D refractive index of these cells is reconstructed using Rytov approximation with projections achieved with an experimental setup utilizing spatial light modulator as described in \cite{lim2019high}. Then, a 2D slice of the refractive index is chosen in the plane of best focus. A total number of 8 cells are used and we rotated and shifted these cells to create a dataset of 136 inhomogeneous cells whose refractive index range is $(1.33,1.41)$. We use 122 of these images for training and 14 for validation. Some examples of HCT-116 refractive index dataset are shown in Fig.~\ref{fig:Fig9}(c).
We freeze the encoder of MaxwellNet and fine-tune its decoder for this new dataset. The training and validation losses are presented in Fig.~\ref{fig:Fig9}(b).  

For 3D MaxwellNet, a dataset of 200 phantoms is created. These 3D phantoms have a spherical shape with some details inside them and the range of their diameter is $1.8-2.4\mathrm{\mu m}$. We randomly choose 180 phantoms for training and 20 phantoms for testing. We train 3D MaxwellNet with the training dataset with batch size of 10. The example of Fig.~\ref{fig:3dMaxwellnet} and Fig.~\ref{fig:tomography3D} is one of the phantoms in the testing dataset.

\begin{figure}[t]
\begin{center}
\begin{tabular}{c}
\includegraphics[height=6.5cm]{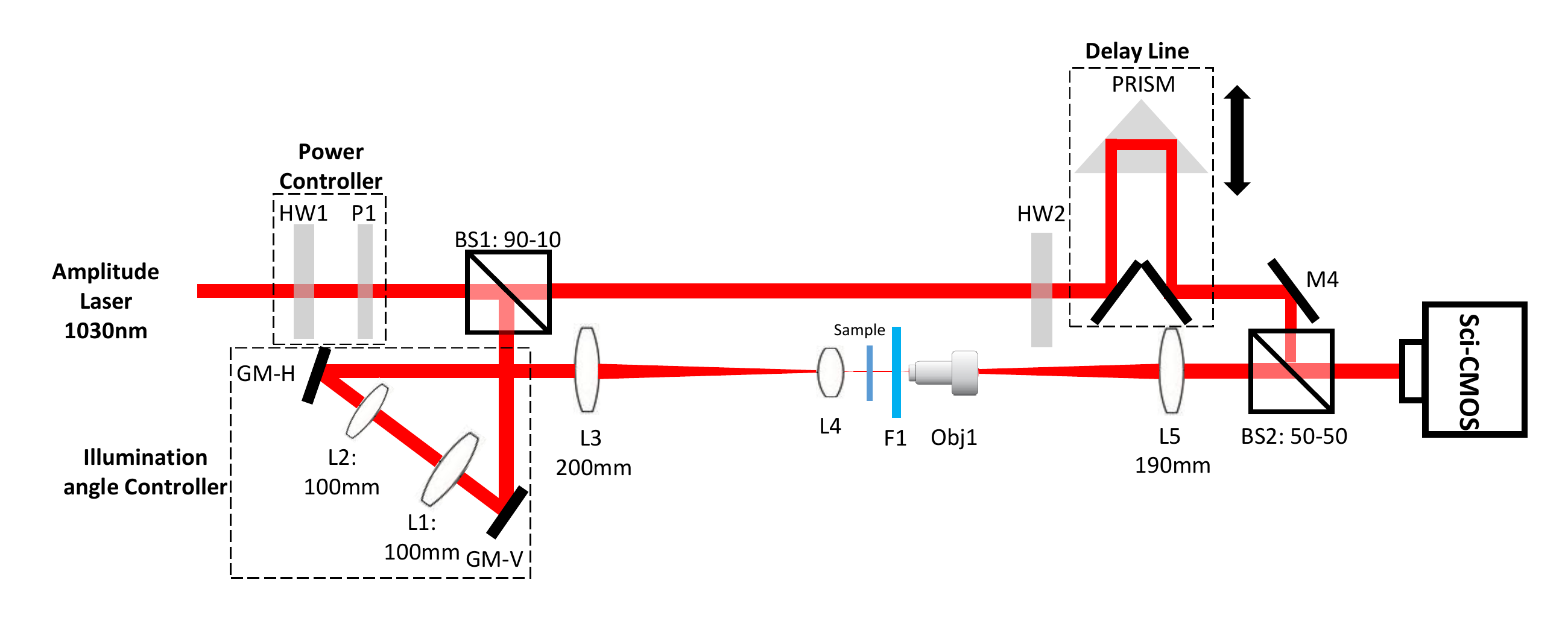}
\end{tabular}
\end{center}
\caption 
{Experimental setup for multiple illumination angle off-axis holography. HW: Half-wave plate, P: Polarizer, BS: Beam splitter, L:Lens, Obj: microscope objective, M: Mirror.} 
\label{fig:setup}
\end{figure}

\section{Experimental Setup for ODT}
\label{sec:appC}
For ODT, we require complex scattered fields from multiple illumination angles. The off-axis holographic setup to accomplish that is shown in Fig.~\ref{fig:setup}. It relies on a ytterbium-doped fiber laser at $\lambda=1.030\mathrm{\mu m}$ whose power is controlled with a half-wave plate and a polarizing beam splitter. The optical beam is divided into the signal and reference arms using a beam splitter (BS1). In the signal arm, we use two galvo mirrors, GM-V and GM-H to control the illumination angle in the vertical and horizontal directions. Using two 4F systems (L1-L4), we image these galvo mirrors on the sample plane, so the position of the beam remains fixed while changing the illumination angle. This way, we can illuminate the sample with a condensed plane wave. The sample is then imaged on the camera (Andor sCMOS Neo 5.5) using another 4F system consisting of a 60X water dipping objective (Obj1) and a tube lens L5. The signal and reference arms are then combined with another beam splitter, BS2 to create the off-axis hologram on the camera. A motorized delay line controls the optical path of the reference arm to match the optical path of the signal arm. 

\subsection{Disclosures}
The authors have no competing interests to declare. 
\subsection{Acknowledgments}
This project was funded by the Swiss National Science Foundation (SNSF) funding number 514481. Additionally, A. Saba would like to thank Dr. Babak Rahmani and Dr. Joowon Lim for their useful comments about the implementation of MaxwellNet and tomographic optimization in Tensorflow. 


\bibliography{report}   
\bibliographystyle{ieeetr}   

\end{spacing}
\end{document}